# Cone-guided fast ignition with *no* imposed magnetic fields

D. Strozzi[a], M. Tabak, D. Larson, M. Marinak, M. Key, L. Divol, A. Kemp, C. Bellei and H. Shay

Lawrence Livermore National Laboratory, Livermore, CA 94550, USA

**Abstract.** Simulations are presented of ignition-scale fast ignition targets with the integrated Zuma-Hydra PIC-hydrodynamic capability. We consider a spherical DT fuel assembly with a carbon cone, and an artificially-collimated fast electron source. We study the role of E and B fields and the fast electron energy spectrum. For mono-energetic 1.5 MeV fast electrons, without E and B fields, ignition can be achieved with fast electron energy $E_f^{ig}$ = 30 kJ. This is 3.5x the minimal deposited ignition energy of 8.7 kJ for our fuel density of 450 g/cm$^3$. Including E and B fields with the resistive Ohm's law $\mathbf{E} = \eta \mathbf{J}_b$ gives $E_f^{ig}$ = 20 kJ, while using the full Ohm's law gives $E_f^{ig}$ > 40 kJ. This is due to magnetic self-guiding in the former case, and $\nabla n \times \nabla T$ magnetic fields in the latter. Using a realistic, quasi two-temperature energy spectrum derived from PIC laser-plasma simulations increases $E_f^{ig}$ to (102, 81, 162) kJ for (no E/B, $\mathbf{E} = \eta \mathbf{J}_b$, full Ohm's law). Such electrons are too energetic to stop in the optimal hot spot depth.

This paper presents work on ignition-scale transport modelling of fast ignition [1] designs. By "transport" we mean the propagation and deposition of fast electrons, with reasonably self-consistent coupling to the background radiation-hydrodynamics. To achieve this we coupled the hybrid-PIC code Zuma [2, 3] to the rad-hydro code Hydra [4]. A detailed publication on this work [3] reports laser-plasma PIC modelling that shows a large fast-electron divergence, along with mitigation ideas based on imposed magnetic fields. Here we discuss the fast-electron energy needed for ignition, $E_f^{ig}$, of an artificially-collimated source. Using the resistive Ohm's law $\mathbf{E} = \eta \mathbf{J}_b$ leads to magnetic self-guiding and reduces $E_f^{ig}$ compared to runs with no E or B fields. However, using a more complete Ohm's law *increases* $E_f^{ig}$ compared to the no-field case. This ordering applies both for a mono-energetic 1.5 MeV and PIC-based fast electron energy spectrum. The latter gives electrons that are too energetic to stop in the optimal hot spot, and raises $E_f^{ig}$ ~3.4x over the 1.5 MeV spectrum.

## 1 Integrated PIC-hydrodynamic modelling with Zuma-Hydra

We summarize Zuma and the Zuma-Hydra coupling here (see [3] for details). Zuma treats the fast electrons by standard relativistic-PIC methods, and injects them according to prescribed distributions. They undergo energy loss and angular scatter following [5] and [6]. The background plasma is treated as a collisional fluid with fixed ions. We eliminate physics on fast (Langmuir or light wave) time scales via reduced equations of motion. In particular, the background current is given by Ampère's law without displacement current: $\mathbf{J}_b = -\mathbf{J}_f + \mu_0^{-1} \nabla \times \mathbf{B}$ where (b, f) denotes (background, fast) current. The magnetic field is evolved by Faraday's law $\partial \mathbf{B}/\partial t = -\nabla \times \mathbf{E}$.

---

[a] strozzi2@llnl.gov

The electric field comes from Ohm's law, the massless force law for background electrons:

$$\mathbf{E} = \mathbf{E}_C + \mathbf{E}_{NC} \qquad \mathbf{E}_C = \boldsymbol{\eta}\cdot\mathbf{J}_b - e^{-1}\boldsymbol{\beta}\cdot\nabla T_{eb} \qquad \mathbf{E}_{NC} = -(en_{eb})^{-1}\nabla p_{eb} - \mathbf{v}_{eb}\times\mathbf{B} . \qquad (1)$$

($\mathbf{E}_C$, $\mathbf{E}_{NC}$) are (collisional, collisionless) terms. $\boldsymbol{\eta}$ and $\boldsymbol{\beta}$ are tensors due to nonzero Hall parameter $\omega_{ce}\tau_e$. We follow "notation II" of Ref. 7 and use their approximate values for $\boldsymbol{\eta}$ and $\boldsymbol{\beta}$. We distinguish between the full Ohm's law Eq. (1) and the resistive Ohm's law $\mathbf{E} = \eta\mathbf{J}_b$ with scalar (unmagnetized) $\eta$. $\eta$ is found following Lee and More [8] with Desjarlais' improvements [9], and the charge state from Ref. 9's modified Thomas-Fermi approach. At the start of a Zuma-Hydra coupling step, Hydra transfers plasma conditions to Zuma. Zuma advances for many of its own timesteps, accumulating energy and momentum deposition rates in each zone. Hydra then runs for many timesteps, using the deposition from Zuma. The cycle then repeats. Both codes were run on Eulerian cylindrical RZ grids.

Fast electrons are injected into Zuma from a distribution factorized into an energy times an angle spectrum. The energy spectrum is either mono-energetic or the PIC-based form from Ref. 3:

$$dN/d\varepsilon = \varepsilon^{-1}\exp[-\varepsilon/\tau_1] + 0.82\exp[-\varepsilon/\tau_2] \qquad \tau_1 = 0.19 \qquad \tau_2 = 1.3 \qquad \text{for } \varepsilon > 0.12. \qquad (2)$$

$\varepsilon = E/T_p$ where $E$ is the kinetic energy and $T_p/m_ec^2 = [1+a_0^2]^{1/2}-1$ is the ponderomotive temperature in the nominal laser intensity, with $a_0^2 = I_0\lambda_0^2/1370$ PW cm$^{-2}$ µm$^2$. The form Eq. (2) fits well the results of a 3D, 360 fs laser-plasma full-PIC simulation with the PSC code [10]. The average $\varepsilon$ in this spectrum is $<\varepsilon> = 1.02$. We scale the spectrum ponderomotively as we vary intensity, though this choice has not yet been validated by PIC runs. The energy spectrum is a key aspect of fast ignition, and is not fully understood. Other work reports a cooler, more favourable scaling [11]. Our PIC data clearly shows two distinct temperatures, which may be important in interpreting experimental data. More details are in Ref. 3. We use $\lambda_0 = 527$ nm, and 52% laser to fast electron power conversion. The fraction of light absorbed by matter in the PSC run was higher, but not all was into fast electrons. Longer PSC runs show a third, high-temperature component with poor fuel coupling. The solid angle spectrum is $dN/d\Omega = \exp[-(\theta/\Delta\theta)^4]$ with $\theta$ the velocity-space polar angle relative to the z axis. $\Delta\theta=10°$ (average $\theta=6.9°$) for our artificially collimated source; the PSC results were matched by the divergent $\Delta\theta=90°$ ($<\theta>=52°$).

## 2 Results with a mono-energetic source

We model an idealized DT fuel assembly with a carbon cone and initial 100 eV temperature, shown in Figure 1. Our peak density $\rho$=450 g/cm$^3$ and $\rho r$=3.0 g/cm$^2$ give an ideal burn-up fraction of 1/3 and fusion yield of 64.4 MJ. Atzeni [12] has found from 2D hydrodynamic simulations with prescribed heating (not self-consistent fast electron dynamics) that the minimal heat deposited in the hot spot needed for ignition is 140 kJ / ($\rho$/100 g/cm$^3$)$^{1.85}$, which for our $\rho$ is 8.7 kJ. The fast electron source in Zuma is injected 20 µm to the left of the cone tip at $z = 0$.

We first find the ignition energy $E_f^{ig}$ for an artificially collimated ($\Delta\theta=10°$), 1.5 MeV mono-energetic source. The optimal DT hot-spot depth [12] is $\rho\Delta z = 1.2$ g/cm$^2$, which removes at most 1.3 MeV from a fast electron. 1.5 MeV electrons mostly stop in this depth. The time pulse is a 19 ps flattop, and intensity profile $I(r) = I_{0f}\exp[-(r/r_{spot})^8]$ with $r_{spot}$=18 µm. Neither of these is optimized for this case, but this $r_{spot}$ gave the smallest $E_f^{ig}$ for $\Delta\theta=10°$, the PIC-based energy spectrum, and the full Ohm's law [3]. Figure 1 displays the yield vs. fast electron energy $E_f$. For no E and B fields, ignition occurs for $E_f^{ig} = 30.4$ kJ. This is 3.5x Atzeni's minimum, due to including a cone, finite beam divergence and angular scattering, and un-optimized time pulse and spot shape. Including fields and the resistive Ohm's law $\mathbf{E} = \eta\mathbf{J}_b$ reduces $E_f^{ig}$ to 20.3 kJ. This is due to magnetic self-guiding by the fast electrons, as shown in the current plots in Figure 2 and magnetic field plots in Figure 3. However, using the full Ohm's law increases $E_f^{ig}$ to > 40 kJ (numerical problems occurred for larger $E_f$). Figure 2 shows less fast current reaching the fuel for this case than the other two.

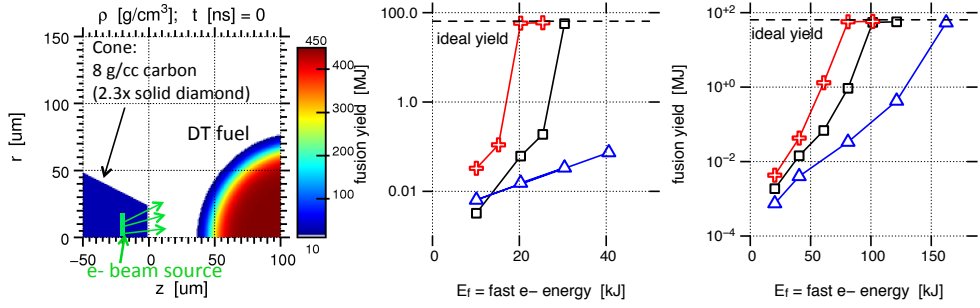

**Fig. 1.** Left: idealized geometry for Zuma-Hydra runs. Middle: fusion yield for $\Delta\theta = 10°$, 1.5 MeV energy spectrum, and no E or B fields (black squares), $\mathbf{E} = \eta \mathbf{J}_b$ Ohm's law (red crosses), and $\mathbf{E}$ from the full Ohm's law Eq. (1) (blue triangles). Right: same as middle but PIC-based energy spectrum Eq. (2).

The effect of E and B fields is similar to Ref. 13: including just the resistive E produces self-guiding of collimated fast electrons, while the full Ohm's law reduces the coupling. In Ref. 13 the full Ohm's law gives better coupling than the no-field case, but we see the opposite. The relative ordering is thus not universal. Our scenario differs from Ref. 13 in the laser pulse, spot shape, and plasma profiles. We have not identified which aspect leads to the different ordering, or simulated their scenario with our codes. The reduced coupling is likely due to the $\nabla n_{eb} \times \nabla T_{eb}$ magnetic field arising from $\mathbf{E} \propto \nabla p_{eb}$. A spherically-symmetric $n_{eb}$ and an azimuthally-symmetric $T_{eb}$ give

$$\partial \mathbf{B}/\partial t = -(en_{eb})^{-1} \nabla n_{eb} \times \nabla T_{eb} \quad \rightarrow \quad \partial B_\phi / \partial t = -(en_{eb} R)^{-1} (dn_{eb}/dR)(\partial T_{eb}/\partial \theta) \ . \tag{3}$$

$\phi$ is the azimuthal angle, and $\theta$ the polar angle with respect to positive $z$. Between the cone tip and the fuel, $dn_{eb}/dR < 0$ while $\partial T_{eb}/\partial \theta > 0$ once heating by fast electrons begins. We therefore generate a positive $B_\phi$, which exerts an outward radial force on fast electrons with $v_z > 0$. Such a $B_\phi$ is seen in Figure 3 for $z$ = 50-70 µm and $r$ < 20 µm. We did not toggle specific non-resistive terms in Ohm's law, e.g. tensor vs. scalar transport coefficients due to Hall parameter $\omega_{ce}\tau_e > 0$. For the Full Ohm's law run with 1.5 MeV source spectrum and $E_f$ = 20 kJ at time 7 ps, $\omega_{ce}\tau_e$ was >1 in most of the transport region and peaked at 13.4. This indicates magnetized transport coefficients should be used.

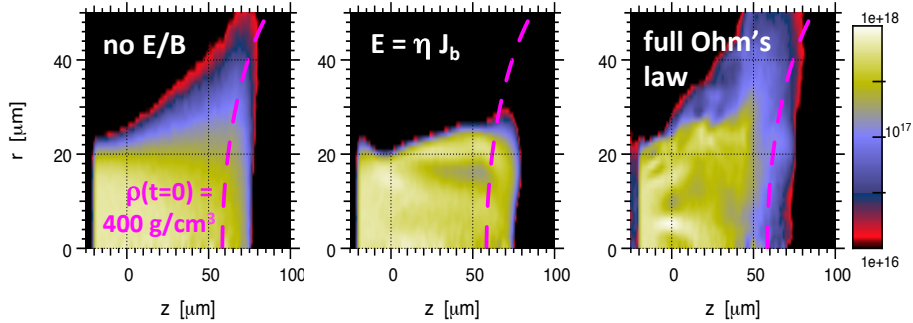

**Fig. 2.** Fast electron current density $|\mathbf{J}_f|$ [A/m$^2$] at $t$ = 10 ps (mid-pulse) for runs with $\Delta\theta = 10°$, 1.5 MeV energy spectrum, and $E_f$ = 20.3 kJ. Left: no E or B fields, middle: $\mathbf{E} = \eta \mathbf{J}_b$ Ohm's law, right: full Ohm's law.

## 4 Results with PIC-based source energy spectrum

Zuma-Hydra runs were performed with the PIC-based energy spectrum Eq. (2), instead of a mono-energetic one (but the same $\Delta\theta = 10°$). Figure 1 reveals a several-fold increase in $E_f^{ig}$ over the mono-energetic cases. The total fast-electron energy is $E_f = ATI_{0f}$, where $A$ and $T$ are the source area and duration. For our time pulse and spot shape, regardless of energy spectrum, $E_f$ [kJ] = $I_{0f} / 5.77 \times 10^{18}$

W/cm$^2$. For our PIC-based spectrum, we have laser intensity $I_0 = = I_{0f}/0.52 = 1.11 E_f \times 10^{19}$ W/cm$^2$ and $<E> \approx 0.783 E_f^{1/2}$ MeV ($E_f$ in kJ). $E_f$ equals the minimal $E_f^{ig} = 8.7$ kJ for $I_0 = 9.5\times10^{19}$ W/cm$^2$, which for the PIC-based source gives $<E>$ = 1.83 MeV. This has significant energy in electrons that do not fully stop in the optimal depth ($E > 1.3$ MeV). To estimate this effect, consider all the fast electrons to have the PIC-based $<E>$, and assume $a_0 \gg 1$. The energy deposited in the ideal depth is

$$E_f^{HS} = A\,T\,[I_{0f}I_{0S}]^{1/2} \qquad I_{0S} = 1.59\times10^{19}\,(\lambda_0/527\text{ nm})^{-2}\text{ W/cm}^2, \qquad (4)$$

or $E_f^{HS} = (2.78\text{ kJ}*E_f)^{1/2}$ (the 2.78 holds for our $A$ and $T$). The Zuma-Hydra runs without E and B fields gave $E_f^{ig} = 102$ kJ, implying $E_f^{HS} = 0.165 E_f = 16.8$ kJ. With a 1.5 MeV spectrum, our runs needed $E_f = 30.4$ kJ to ignite, which from stopping suggests $E_f^{HS}=(1.3/1.5)E_f = 26.3$ kJ (neglecting angular divergence and scatter). The PIC-based spectrum needed 102/30.4 = 3.4 times the fast electron energy of a 1.5 MeV spectrum to ignite. This is a substantial increase but *less* than the factor $26.3^2/(2.78*30.4) = 8.2$ our naïve stopping model requires to deposit the same 26.3 kJ. This may stem from increased stopping of the cold part of the PIC source (our estimate assumes all electrons have the average energy), or more angular scatter in the 1.5 MeV case.

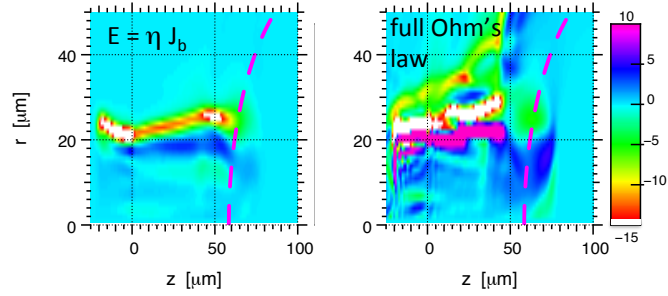

**Fig. 3.** Azimuthal field $B_\phi$ (capped) [MG] for two runs from Figure 2. Left: $\mathbf{E}=\eta\mathbf{J}_b$, right: full Ohm's law.

Our work suggests 300 kJ of short-pulse laser is adequate for fast ignition - if the divergence problem is solved. This can be reduced by approaches to cool the source spectrum: LPI that produces an intrinsically cooler spectrum (e.g. shorter laser wavelength), targets with larger spot sizes or fuel dimensions (lower density), and longer laser pulses; or by enhanced electron stopping (micro-instabilities, orbit roll-up by magnetic fields).

This work was performed by LLNL under U.S. DoE Contract DE-AC52-07NA27344, and partly supported by LDRD project 11-SI-002 and the Office of Fusion Energy Sciences.

## References

1. M. Tabak, J. Hammer, M. Glinsky, et al., Phys. Plasmas **1**, 1626 (1994)
2. D. Larson, M. Tabak, T. Ma, Bull. Am. Phys. Soc. **55**, 15 (2010)
3. D. J. Strozzi, M. Tabak, D. J. Larson, et al., Phys. Plasmas **19**, 072711 (2012)
4. M. M. Marinak, G. D. Kerbel, N. A. Gentile, et al., Phys. Plasmas **8**, 2275 (2001)
5. A. A. Solodov, R. Betti, Phys. Plasmas **15**, 042707 (2008)
6. S. Atzeni, A. Schiavi, J. R. Davies, Plasma Phys. Controlled Fusion **51**, 015016 (2009)
7. E. M. Epperlein, M. G. Haines, Phys. Fluids **29**, 1029 (1986)
8. Y. T. Lee, R. M. More, Phys. Fluids **27**, 1273 (1984)
9. M. P. Desjarlais, Contrib. Plasma Phys. **41**, 267 (2001)
10. A. J. Kemp, B. I. Cohen, L. Divol, Phys. Plasmas **17**, 056702 (2010)
11. P. A. Norreys, R. H. H. Scott, K. L. Lancaster, et al., Nucl. Fusion **49**, 104023 (2009)
12. S. Atzeni, A. Schiavi, C. Bellei, Phys. Plasmas **14**, 052702 (2007)
13. Ph. Nicolaï, J.-L. Feugeas, C. Regan, et al., Phys. Rev. E **84**, 016402 (2011)